# Gold nanorod radioenhancement: interpreting Monte Carlo dose enhancement across geometry, composition, coating, and localization

Ali Taheri [a,*], Hans Rabus [b], Mayeen Uddin Khandaker [a,c,d], Farhad Moradi [e], David Andrew Bradley [a,f]

[a] Applied Physics and Radiation Technologies Group, CCDCU, Faculty of Engineering and Technology, Sunway University, 47500 Bandar Sunway, Selangor, Malaysia
[b] Physikalisch-Technische Bundesanstalt (PTB), 10587 Berlin, Germany
[c] The Miyan Research Institute, International University of Business Agriculture and Technology, 4, Embankment Drive Road, Sector-10, Uttara Model Town, Dhaka-1230, Bangladesh
[d] Department of Physics, College of Science, Korea University, 145 Anam-ro, Seongbuk-gu, Seoul, 02841, Republic of Korea
[e] Department of Biomedical Imaging, Faculty of Medicine, Universiti Malaya, Kuala Lumpur, Malaysia
[f] School of Mathematics and Physics, University of Surrey, Guildford, GU27XH, United Kingdom

* Corresponding author email: alitmedical@gmail.com, alitaheri@sunway.edu.my

## Abstract

Nanoparticle radiosensitization depends on multiple interconnected properties of the nanoparticle and its spatial distribution, yet geometry, elemental composition, surface coating, and localization are commonly investigated separately, and the resulting biological effect is rarely addressed within the same framework. Drawing on a series of TOPAS Monte Carlo studies conducted as part of a PhD research project, we integrate four investigations of metallic nanorods, primarily gold nanorods (AuNRs), that examine these parameters within closely related investigations. The combined results show that each parameter has a different characteristic range of influence. Nanorod geometry primarily modifies the spectrum and distribution of emitted low-energy Auger-Meitner electrons within tens of nanometers of the surface. The elemental composition of the nanoparticle produces substantial differences in near-field dose enhancement that diminish rapidly with distance and become negligible beyond approximately 1 µm. Coating layers reduce the fluence of low-energy secondary electrons (below 3.5 keV), which are absorbed within roughly 150 nm of the surface, and lower the physical dose contribution by 1 to 7%. Because coatings mainly remove low-energy electrons, the radiolytic yield changes through a shift in the electron energy spectrum and does not follow the proximal dose response directly. At the vascular scale, the region of enhanced dose extends to approximately 10 µm from the microvessel wall, so nanoparticle localization determines where the biological target lies relative to the sources. Together, these findings indicate that no single design parameter is a universal predictor of radiosensitization, and that translating physical radioenhancement into biological effect depends on whether the target sits within the range over which each parameter acts. We further show that charged-particle equilibrium (CPE) correction, applied where beam confinement required it, substantially changes calculated dose enhancement ratios. Future Monte Carlo studies should therefore report scoring distance and CPE-corrected values explicitly to make results comparable across studies.

# 1 Introduction

Gold nanoparticles have long been investigated as radiation enhancers, since interactions of X-rays with high-Z materials generate additional electrons that increase local energy deposition [1]–[3]. Monte Carlo track-structure simulations have since provided a means of resolving these interactions from the nanoparticle scale to cellular dimensions, including the transport of low-energy secondary and Auger–Meitner electrons [4]–[6]. Throughout this Letter, "radioenhancement" denotes the physical increase in energy deposition and secondary electron emission caused by the nanoparticle. "Radiosensitization" is the overall effect, which comprises this physical component and the resulting radiochemical and biological effects, such as radiolytic yield, DNA damage, cell kill, and tumor control.

A recurring difficulty in interpreting the literature is that numerical values for radioenhancement depend strongly on how and where dose is scored. A large DER close to an isolated nanoparticle does not represent the same quantity as the dose enhancement averaged over a cell, vessel, or larger tissue volume, and confined irradiation geometries can further violate charged-particle equilibrium (CPE), producing substantial differences in calculated DER unless the missing contribution from charged particles generated outside the scoring region is properly addressed [7]–[9]. Consequently, numerical DER values obtained at different spatial scales or under different beam and equilibrium conditions should not be compared directly, and the same caution applies before any radioenhancement result is read as a radiosensitization outcome.

During our research on gold nanorod (AuNR) radioenhancement, four simulation studies utilizing TOPAS code [10] and its TOPAS-nBio extension [11], [12] examined complementary aspects of nanoparticle design: geometry and orientation [13], elemental composition [14], coating layers [15], and vascular localization [16]. Each study addressed a specific question in isolation. Synthesized together, they provide a broader picture of how these parameters influence different stages of the interaction between radiation, nanoparticle, surrounding medium, and biological target. The purpose of this Letter is to make that connection explicit.

# 2 Nanoparticle characteristics and localization

## 2.1 Nanorod geometry and orientation

The first study examined how AuNR geometry and orientation affect radiation-induced electron emission and dose enhancement [13]. When a gold nanorod and a gold nanosphere (AuNP) of comparable volume and cross-sectional area were irradiated under kilovoltage X-rays, their microscale dose enhancement ratios (DERs) differed by only several percent. In contrast, their emitted electron spectra differed more substantially, particularly for M-shell Auger-Meitner and other low-energy electrons (with energies below 3.5 keV), with changes of approximately 13-16% observed. A stronger dependence was obtained when the surface-area-to-volume ratio (SA:V) of the nanorods was varied at constant volume. Increasing SA:V produced an approximately 35% reduction in the low-energy electron fluence ratio (emitted electron fluence normalized to the photon fluence inside the nanorod), while the total electron number ratio (total emitted electrons per photon fluence inside the nanorod) and microscale DER showed much smaller changes. Orientation of the nanorod relative to the radiation beam also altered the electron fluence and DER, although the effect was smaller than that associated with the major geometry changes. These results indicate that nanoparticle geometry can substantially modify the local electron field without producing a correspondingly large change in dose averaged over a larger volume. The distinction is important, as changes in the electron population close to the nanoparticle should not automatically be interpreted as equivalent changes in cellular or tissue-level radioenhancement. The comparable microscale DER between AuNRs and AuNPs under these conditions indicates that nanorods represent a physically viable alternative to the conventionally studied nanospheres, without requiring a compensating change in dose enhancement.

### 2.2 Elemental composition

Upon the establishment of comparability of the radioenhancement effect of rod-shaped NPs with sphere-shaped NPs, the influence of elemental composition was examined by comparing Au, Pt, Hf, Gd and Ce nanorods with the same dimensions [14]. The calculated local dose enhancement showed a strong dependence on material in the immediate vicinity of the nanorod. At approximately 130 nm from the nanorod centre, Au and Pt produced local DERs of approximately 6.5–7, compared with ~4 for Hf and ~2 for Gd and Ce under the conditions studied. The difference between materials decreased rapidly with distance: beyond approximately 250–300 nm, the material-dependent differences fell below the 4% statistical significance threshold adopted in that study, and beyond approximately 1 μm the local DER itself approached unity for all five materials, leaving no meaningful enhancement for material choice to differentiate. Thus, elemental composition has a pronounced effect when dose is evaluated sufficiently close to the nanoparticle, but its influence becomes increasingly less important as the scoring volume extends beyond the region where the additional low-energy electron contribution is concentrated. This finding also provides an important qualification to statements that a particular high-Z material is intrinsically a more effective radiosensitizer. The answer depends on the radiation quality, nanoparticle concentration, geometry, and, critically, the spatial volume over which the enhancement is evaluated.

### 2.3 Surface coating

The study continued with the role of surface functionalization, considering silica, polyethylene glycol (PEG), polyethyleneimine (PEI), and cetyltrimethylammonium bromide (CTAB) coating layers with thicknesses of 25 and 50 nm surrounding AuNRs [15]. Introducing the coating layer reduced the calculated physical dose contribution in the water surrounding the nanoparticle (dose in a 5 μm water cube outside the nanorod and coating, normalized to the incident photon fluence) by approximately 1–7%, with thicker coatings generally producing greater attenuation. Differences between the coating materials themselves were comparatively small. The chemical stage, however, produced a more complicated picture. The calculated radiolytic yields did not simply follow the reduction in physical dose. For example, increasing silica thickness produced a modest increase in the calculated total G-value (the number of molecules formed per 100 eV of absorbed energy, summed over the seven radiolytic species scored) under the conditions studied. This illustrates why physical dose enhancement and radiolytic yield should not be treated as interchangeable measures of radiosensitization. The chemical simulations in that study did not explicitly represent coating-specific interfacial chemistry or radical scavenging reactions, so these results establish the influence of the coating on simulated electron transport and the bulk radiolytic environment, but not that different coating chemistries produce distinct biological radiosensitization through specific interfacial reactions, a knowledge gap that requires further investigation.

## 2.4 Vascular localization

The study moved from nanoparticle-scale properties to the spatial distribution of AuNRs within a model microvessel [16]. Three distributions were compared: homogeneous dispersion within the vessel lumen, attachment to the inner vessel wall, and attachment to the outer wall. The simulations used cell-resolved endothelial scoring and explicitly corrected the DERs for the lack of CPE associated with the confined microscopic beam. At a reference concentration of 18 mg Au/g water, inner-wall-attached AuNRs produced microvessel-averaged DERs of approximately 2.0 at 50 kVp and 1.7 at 100 kVp, with values between 1.4 and 2.2 in individual endothelial cells. At 50 mg Au/g water and 50 kVp, the microvessel-averaged DER reached approximately 3.8, in accordance with the observed linear increase in DER with NP concentration. An analytical estimate of the dose contribution from AuNRs in adjacent vessel segments increased the endothelial DERs by approximately 5–10%, without changing the relative trends between localization patterns. The radial dose profiles showed that the physical enhancement decreased rapidly with distance from the vessel wall. Beyond approximately 10 µm, the local DER fell below ~1.05 under the conditions investigated. Thus, the electron emission of an individual nanorod is the same in all three distributions. Localization changes the distance between the nanorods and the endothelial cells, and therefore how much of the emitted electron fluence reaches the cells and how the enhanced-dose region is positioned relative to the biological structure.

## 2.5 Synthesis across the four studies

Taken together, the four studies suggest that nanoparticle radioenhancement is not ruled by a single nanoparticle property or adequately described by a single dose-enhancement value. The effect observed at any given scale depends on how the nanoparticle, its surrounding environment, and the biological target are spatially related.

*Table 1 Primary influence, dominant spatial regime, and key observation for each investigated parameter across the four studies.*

| Design parameter | Primary influence | Dominant spatial range | Key observation from the four studies |
|---|---|---|---|
| Geometry | Electron emission and spatial distribution | ≲ 200 nm | Substantial changes in low-energy electron fluence can occur with relatively small changes in microscale DER |
| Composition | Magnitude of electron-mediated enhancement | ≲ 300 nm, negligible beyond ≈ 1 µm | Material-dependent differences diminish rapidly with distance |
| Coating | Electron transport and radiolytic environment | ≲150 nm from the nanorod surface (physical), extending across the $(5\ \mu m)^3$ water volume in the chemical stage | Physical attenuation is modest, while coating-specific chemical effects remain incompletely resolved. Coatings were studied on gold cores only, so core-coating interactions are untested. |
| Localization | Spatial overlap of enhancement with the target | ≈ 10 µm from the vessel wall | The position of the nanoparticle can determine how much of the local enhancement is received by the target cells |

The important point is therefore that these four parameters neither act independently nor occupy separate spatial regimes. Each one changes a different link in the chain that sets how radiation energy is distributed around the nanoparticle: electron emission (geometry and composition), electron transport (coating), and the position of the target relative to the sources (localization). The local DER falls off steeply with distance from the nanoparticle, and a volume-averaged DER over a cell or vessel combines this near-field enhancement with the unenhanced remainder of the volume. The biological relevance of the enhancement ultimately depends on whether the affected region overlaps the structure of interest.

This has a direct consequence for how nanoparticle radiosensitization should be compared across studies. A high local DER calculated close to an isolated nanoparticle and a moderate volume-averaged DER over a cell or vessel are not competing measurements of the same quantity. Compilations that pool reported dose enhancement values into a single range, such as that of Li et al. [17], therefore combine values scored in different volumes and with and without CPE correction. The spread of such a range reflects methodology as much as nanoparticle physics, and it cannot be used to link physical enhancement to biological effect unless the scoring volume and equilibrium conditions are stated for each value.

Considered together, the four studies therefore point toward a more useful design question: not simply which nanoparticle produces the highest radioenhancement, but which nanoparticle configuration produces the relevant enhancement at the spatial location and biological scale of the intended target. This provides a common framework for interpreting the individual findings without assuming that one design parameter universally dominates the others.

## 3 Scope, limitations, and implications for future Monte Carlo studies

Connecting these physical and radiochemical findings to biological endpoints such as DNA damage or cell survival is the natural next step, and one that remains only partially closed here. For the vascular localization results specifically, single- and double-strand break counts were computed for each localization scenario as part of the broader PhD programme and tracked the DER trends closely, yet this analysis is not part of the manuscript cited here [16], and is aimed to be prepared for separate submission. The biological consequence of a given local dose distribution will in any case depend on nanoparticle uptake, cellular architecture, oxygenation, DNA organization, and repair processes, factors outside the scope of the present physical and radiochemical comparison. Connecting the dose distributions reported here to validated biological endpoints, using consistent transport, chemical, and biological assumptions, would determine whether differences pronounced within a few hundred nanometres remain important after averaging over nuclei or cells.

The four studies also point to a narrower methodological gap than is sometimes assumed. Nanoparticle concentration, spatial distribution, and photon spectrum are already routinely reported in the literature. What is inconsistently reported, and what has been shown to substantially affect calculated DER, is the treatment of charged-particle equilibrium and the explicit link between the scoring distance chosen and the biological target the study is meant to inform [7]–[9]. A large DER scored close to a nanoparticle under a confined beam and a lower DER scored across a microvessel under CPE correction describe different physical quantities, not competing estimates of one quantity. Reporting a DER without its CPE treatment and scoring geometry leaves the number impossible to interpret.

## 4 Conclusion

Considered separately, the four studies each answer a narrow question about radioenhancement, the physical increase in energy deposition or radiolytic yield, whether geometry and elemental composition alter it, whether coating suppresses it as much as generally assumed, and how vascular localization redistributes it spatially. Considered together, however, they answer a different question of at what distance from the nanoparticle does a given design parameter stop contributing to radioenhancement. This has a direct bearing on radiosensitization, the biological effect is meant to produce, since a design change confined to tens of nanometres from the nanorod surface can only translate into a noticeable biological effect if the sensitive target sits within that same range. Radiosensitization efficiency is therefore not a property of the nanoparticle alone, but of the match between the nanoparticle's characteristic length scale and the scale of the biological target the enhancement is meant to reach.

**Acknowledgement**:

Sunway University PhD Studentship support received by A. T. is acknowledged. The authors also thank Dr. José Ramos-Méndez for valuable insight and advice on TOPAS-nBio simulations.